\documentclass{article}
\usepackage{graphicx}
\usepackage[a4paper]{geometry}
\usepackage{amsmath}
\usepackage{amssymb}
\usepackage[numbers]{natbib}
\usepackage{setspace}
\usepackage{authblk}
\usepackage{booktabs}
\usepackage[bookmarks=False]{hyperref}
\hypersetup{
	colorlinks=true,
	linkcolor=blue,
	filecolor=magenta,
	urlcolor=cyan,
}

\newcommand{\prob}{\mathcal{P}}
\newcommand{\ent}{\mathcal{H}}
\newcommand{\oinfo}{\mathcal{O}}
\newcommand{\tc}{\mathcal{TC}}
\newcommand{\dtc}{\mathcal{DTC}}
\newcommand{\mi}{\mathcal{I}}
\newcommand{\sinfo}{\mathcal{S}}

\title{The topology of synergy: linking topological and information-theoretic approaches to higher-order interactions in complex systems}
\date{\today}
\author[1,2]{Thomas F. Varley} 
\author[3,4]{Pedro A.M. Mediano}
\author[1,5]{Alice Patania}
\author[1,2]{Josh Bongard}

\affil[1]{Vermont Complex Systems Institute, University of Vermont, Burlington, VT, USA}
\affil[2]{Department of Computer Science, University of Vermont, Burlington, VT, USA}
\affil[3]{Department of Computing, Imperial College London, London, UK}
\affil[4]{Division of Psychology and Language Sciences, University College London, London, UK}
\affil[5]{Department of Mathematics, University of Vermont, Burlington, VT, USA}

\begin{document}

\maketitle

\begin{abstract}
    The study of irreducible higher-order interactions has become a core topic of study in complex systems, as they provide a formal scaffold around which to build a quantitative understanding of emergence and emergent properties. Two of the most well-developed frameworks, topological data analysis and multivariate information theory, aim to provide formal tools for identifying higher-order interactions in empirical data. Despite similar aims, however, these two approaches are built on markedly different mathematical foundations and have been developed largely in parallel - with limited interdisciplinary cross-talk between them. In this study, we present a head-to-head comparison of topological data analysis and information-theoretic approaches to describing higher-order interactions in multivariate data; with the aim of assessing the similarities and differences between how the frameworks define ``higher-order structures." We begin with toy examples with known topologies (spheres, toroids, planes, and knots), before turning to more complex, naturalistic data: fMRI signals collected from the human brain. We find that intrinsic, higher-order synergistic information is associated with three-dimensional cavities in an embedded point cloud: shapes such as spheres and hollow toroids are synergy-dominated, regardless of how the data is rotated. In fMRI data, we find strong correlations between synergistic information and both the number and size of three-dimensional cavities. Furthermore, we find that dimensionality reduction techniques such as PCA preferentially represent higher-order redundancies, and largely fail to preserve both higher-order information and topological structure, suggesting that common manifold-based approaches to studying high-dimensional data are systematically failing to identify important features of the data. These results point towards the possibility of developing a rich theory of higher-order interactions that spans topological and information-theoretic approaches while simultaneously highlighting the profound limitations of more conventional methods. 
\end{abstract}

\onehalfspacing
\section{Introduction}

Complex systems, such as brains, societies, and ecosystems are defined by emergent coordination between large numbers of distinct elements (eg. neurons, individuals, or species). Consequently, a core problem in the study of any complex system is understanding the structure of the interactions between elements and the relationship between the ``parts" and the ``whole" (a philosophical tradition known as mereology \cite{cotnoir_what_2021}). Since the inception of complexity science, one of the most successful models that has been developed is that of the network \cite{barabasi_network_2016,sporns_networks_2010}. In a network, the basic objects of study are 1) elements (sometimes referred to as nodes or vertices) and 2) dyadic interactions between elements (typically called edges or links). The dyadic nature of interactions is crucial: in a network, the only directly accessible dependencies are between pairs of elements. Interactions between multiple elements (such as meso-scale communities \cite{betzel_community_2020,traag_community_2009}, multi-element motifs \cite{sporns_motifs_2004}, etc) must be built from combinations of lower-order (pairwise) dependencies. In many networks, this limitation is not problematic: for example, in an airline network, planes have defined origins and destinations. In those cases, having a structure composed of a dyadic building block is very natural. However, in statistical networks (also called ``functional connectivity" networks), the pairwise restriction is not totally natural: it is possible to ask about the dependency that is ``irreducibly" intrinsic to sets of three or more elements \cite{rosas_disentangling_2022}. A classic example of this is the exclusive-OR (XOR) gate, which forms the basis of modern cryptography. In a trivariate XOR gate, there is no correlation between any pair of variables, but the entire triad contains one bit of information (for details, see Supplementary Material 2).

Recent analyses, both formal and empirical, have suggested that in statistical networks, meso-scale structures constructed from pairwise dependencies (eg. Pearson correlation or mutual information) are systematically biased in ways that fail to represent genuine higher-order interactions that are ``greater than the sum of their parts" \cite{varley_partial_2023,varley_multivariate_2023}. There remains a need for well-developed mathematical and statistical frameworks to directly assess higher-order relationships in complex systems. Several different approaches to assessing higher-order interactions in data have been proposed in recent years \cite{battiston_networks_2020}, and here we discuss two of the most well-developed: topological data analysis (TDA) \cite{chazal_introduction_2017,gholizadeh_short_2018} and multivariate information theory \cite{rosas_quantifying_2019,williams_nonnegative_2010,varley_information_2024}. Both formalisms provide frameworks by which higher-order interactions between three or more variables in a dataset can be identified, however, these branches of applied mathematics have developed almost entirely in parallel, with limited cross-fertilization between them. As such, it remains unclear to what extent ``higher-order" interactions in the topological sense reflect the same kind of ``higher-order interaction" in the information-theoretic sense.

A wrinkle in the problem of comparing topological and information-theoretic approaches to higher-order interaction is that the topological approach only defines one ``kind" of higher-order interaction (based on the dimensionality of structures in the data manifold -- cycles, voids, connected components, etc), while the information-theoretic approach often defines two distinct ``kinds" of higher-order interactions: redundancy and synergy. Redundancy constitutes information that is duplicated over multiple proper subsets of variables simultaneously: for example, given a set of random variables $\{X_1,X_2,\ldots,X_k\}$ the redundant information is that information that could be learned by observing $X_1$ alone or $X_2$ alone or $X_3$ alone and so on. In contrast, synergy constitutes information that is present only in the joint state of multiple variables and cannot be learned by observing any proper subset, i.e. synergy is that information that can \textit{only} be learned when $X_1$ and $X_2$ and $X_3$ and so on are observed together.

This study is broken into two parts: in the first part, we apply measures of higher-order information to point clouds sampled from manifolds with diverse, known topologies (spheres, toroids, planes, knots, etc). This lets us build intuition by comparing and contrasting simple, easy-to-understand cases. In the second part, we apply measures from topological data analysis and multivariate information theory to fMRI data recorded from the human brain. The study of higher-order interactions has been particularly popular in neuroscience, as the question of how many interacting ``parts" produce a coherent, emergent ``whole" is a central question in the field. Both topological and information-theoretic approaches have found empirical links been higher-order interactions and diverse neural phenomenon including consciousness \cite{petri_homological_2014,luppi_synergistic_2024,luppi_reduced_2023,varley_topological_2021}, cognitive performance \cite{santoro_higher-order_2024,anderson_topological_2018,varley_information-processing_2023}, aging \cite{gatica_high-order_2021}, neurodegeneration \cite{herzog_genuine_2022,rutkowski_mild_2024,xu_topology-based_2024}, structural and genomic data \cite{sizemore_importance_2019}, and more. By directly comparing topological and information-theoretic approaches in complex, empirical data, we aim to deepen our understanding of higher-order statistics in naturalistic circumstances.

We now turn to providing brief introductions to the basic machinery of multivariate information and topological data analysis, with a specific focus on how they represent higher-order interactions in data. 

\subsection{Multivariate information theory and higher-order interactions}

For an $N$-dimensional random variable $\textbf{X}=(X_1, ..., X_N)$, which takes values \textbf{x} from the support set $\boldsymbol{\mathfrak{X}}$ according to the probability distribution $\prob(\textbf{X})$, the Shannon entropy of \textbf{X} is defined as:
\begin{align}
    \ent(\textbf{X}) = -\sum_{\textbf{x}\in{\boldsymbol{\mathfrak{X}}}}\prob(\textbf{x})\log_2\prob(\textbf{x}).
\end{align}
The entropy can be understood as quantifying an observer's uncertainty about the state of \textbf{X}. While entropy is commonly presented as a measure of ``information", it is better understood as a measure of uncertainty, and information arises from the \textit{reduction} in uncertainty. We can confirm this intuition by considering the simplest measure of information: the bivariate mutual information:

\begin{align}
    \mi(X_1;X_2) &= \ent(X_1) - \ent(X_1|X_2) \\
    &= \ent(X_2) - \ent(X_2|X_1)
\end{align}

The information that random variable $X_1$ discloses about another random variable $X_2$ (the mutual information) is the difference between our initial uncertainty about the state of $X_1$ ($H(X_1)$) and the uncertainty that remains after learning the state of $X_2$ (the conditional entropy: $H(X_1|X_2)$). The mutual information is a symmetric measure that shows how ``information" is associated with the reduction in uncertainty. 

To go beyond the pairwise case and consider multiple interacting elements simultaneously, several different measures have been introduced, each of which is sensitive to different notions of ``structure": the total correlation, the dual total correlation, the O-information, and the S-information. See Rosas et al.~\cite{rosas_characterising_2024} for a discussion of the relationship between these measures.

The total correlation \cite{watanabe_information_1960} (also independently derived as the ``integration" \cite{tononi_measure_1994}) measures the degree to which a multivariate random variable deviates from independence: 

\begin{align}
        \tc(\textbf{X}) &= \sum_{i=1}^{N} \ent(X_i) - \ent(\textbf{X}). 
\end{align}

If all variables in \textbf{X} are independent, then $\tc(\textbf{X})=0$ bit. Conversely, if all $X_i$ are deterministic functions of each-other, then the total correlation achieves its maximum possible value of $\ent(\textbf{X})-1$ bit. In the case of two variables the total correlation reduces to the classic, bivariate Shannon mutual information: $\tc(X_1;X_2)=\mi(X_1;X_2)$. Rosas et al., described the total correlation as a measure of the ``collective constraints" imposed on the whole system: the more constrained the joint distribution, the higher the total correlation \cite{rosas_quantifying_2019}.

The second measure of higher-order information-sharing is the dual total correlation \cite{abdallah_measure_2012}:
\begin{align}
    \dtc(\textbf{X}) = \ent(\textbf{X}) - \sum_{i=1}^{N}\ent(X_i|\textbf{X}^{-i}).
\end{align}
The dual total correlation quantifies all that information that is shared by two or more variables, and has become an increasingly popular measure for genuine higher-order interactions in neuroscience \cite{herzog_genuine_2022,li_higher-order_2023,varley_multivariate_2023}. Like the total correlation, in the case of two variables the total correlation reduces to the classic, bivariate Shannon entropy: $\dtc(X_1;X_2)=\mi(X_1;X_2)$, but it's limit behavior is more complex than the total correlation. Like the total correlation, it is zero when all elements are independent, but it is also low (but non-zero) in the case of total synchrony as well: instead it is maximized when all elements are integrated by complex, multipartite dependencies \cite{abdallah_measure_2012}. Rosas et al., described the dual total correlation as quantifying the amount of ``shared" information present in the system \cite{rosas_quantifying_2019}, highlighting how different measures of higher-order information quantify different notions of structure. 

The third measure is the O-information. First introduced by James and Crutchfield as the ``enigmatic information" \cite{james_anatomy_2011} and latter re-examined and renamed by Rosas et al., \cite{rosas_quantifying_2019}, the O-information is the difference between the total correlation and the dual total correlation:
\begin{align}
    \oinfo(\textbf{X}) = \tc(\textbf{X}) - \dtc(\textbf{X}).
\end{align}
The O-information is a signed measure that quantifies the overall balance of higher-order redundancy and synergy in the system. If $\oinfo(\textbf{X}) > 0$, then the system is redundancy-dominated, while if $\oinfo(\textbf{X})<0$, then the system is synergy-dominated.  In the specific case of three variables (which is what is used exclusively in this manuscript), the O-information can be written out as a whole-minus-sum measure:
\begin{align}
    \label{eq:trivariate_o}
    \oinfo(X_1,X_2,X_3) = -1\times\big(\tc(X_1;X_2;X_3) - \sum_{i<j}\mi(X_i;X_j)\big).
\end{align}
The final measure is the S-information (first introduced by James et al., using the tongue-in-cheek ``very mutual information" \cite{james_anatomy_2011}) and later explored by Rosas et al. \cite{rosas_quantifying_2019}:
\begin{align}
    \label{eq:s_info}
    \sinfo(\textbf{X}) &= \tc(\textbf{X}) + \dtc(\textbf{X}) \\ 
    &= \sum_{i=1}^{N}\mi(X_i;\textbf{X}^{-i})
\end{align}
The S-information is equal to the sum of the mutual information between each element and the rest of the system. In a sense it quantifies the total amount of information in the structure of the system (although note that the S-information can actually be greater than the entropy, since redundant information gets double-counted). 

Comparing different systems with the O-information is non-trivially difficult, and most analyses typically just consider the sign (whether the system is redundancy- or synergy-dominated). The absolute value of the O-information depends on the total deviation from independence, which makes it difficult to compare different systems with different values of $\tc(\textbf{X})$ and $\dtc(\textbf{X})$ (see Liardi et al.~\cite{liardi2024null} for a discussion on normalizing information quantities). One possible way to correct for this is by normalizing the O-information:
\begin{align}
\label{eq:o_hat}
    \bar{\oinfo}(\textbf{X}) &= \oinfo(\textbf{X}) / \sinfo(\textbf{X}) \\
    &= \frac{\tc(\textbf{X})-\dtc(\textbf{X})}{\tc(\textbf{X}) + \dtc(\textbf{X})}
\end{align}
The normalized $\bar{\oinfo}(\textbf{X})$ is bounded by the range [-1,1] (although the bound is not tight), and since the S-information is strictly non-negative, the sign retains the same interpretation. Normalizing by the S-information controls for the variable total amounts of information in the higher-order structure, making direct comparison between different systems more precise.  

\subsubsection{K-nearest neighbors based information estimators}
\label{sec:knn}

\begin{figure}
    \centering
    \includegraphics[width=1\linewidth]{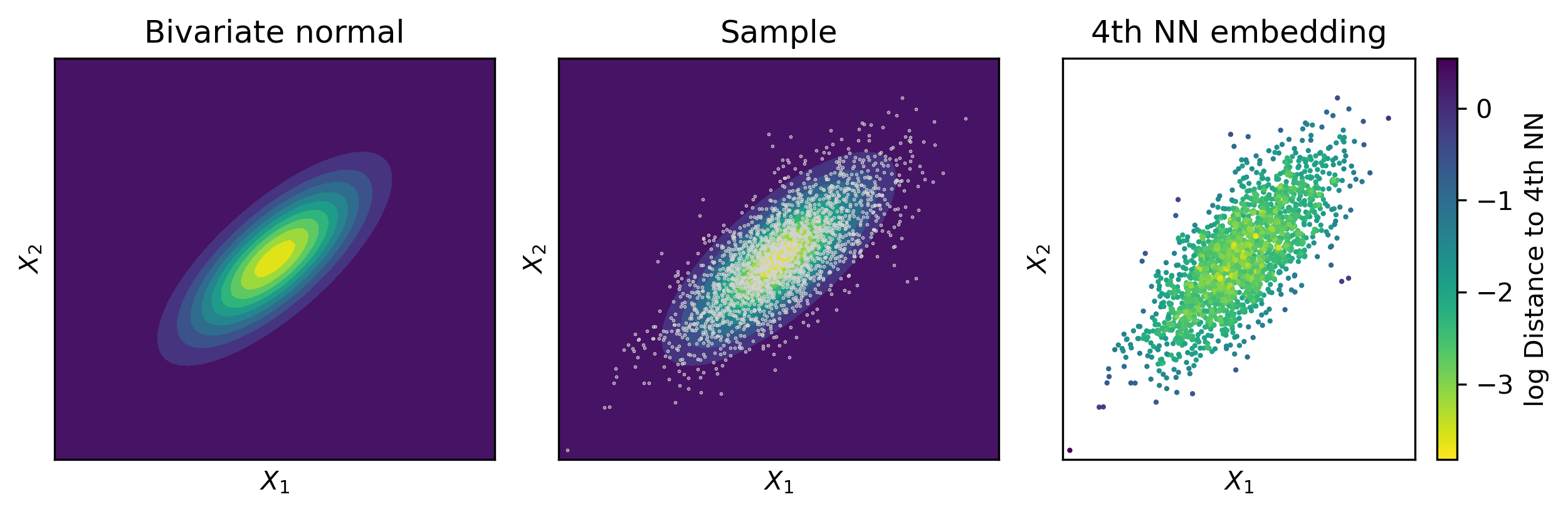}
    \caption{\textbf{Estimating underlying distributions via K-nearest neighbors measures.} This cartoon demonstrates how discrete K-nearest neighbors analyses can be used to estimate the structure of a continuous, underlying distribution. Consider a bivariate normal distribution (left): if we sample a large number of points from it (center), we see that the density of the point cloud tracks the underlying local probability density around each point. If we then compute the distance to the fourth nearest neighbor, and color the points (right), we see how the the distribution of distances roughly recapitulates the underlying bivariate Gaussian.}
    \label{fig:koz}
\end{figure}

\begin{figure}
    \centering
    \includegraphics[width=\linewidth]{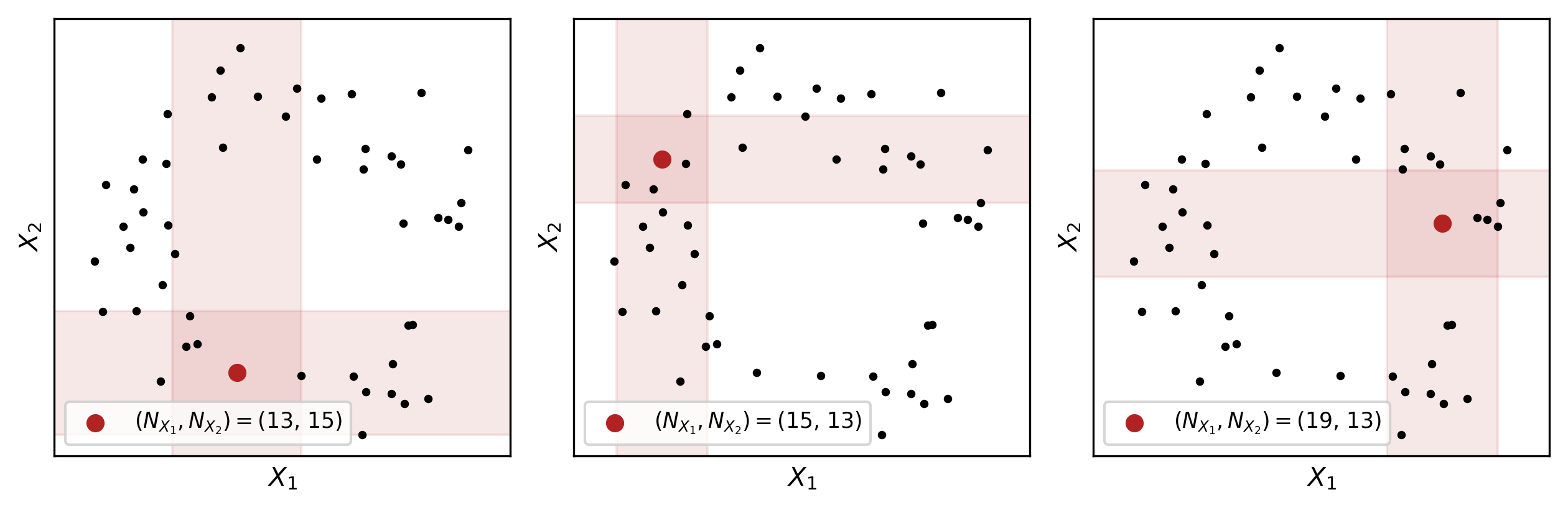}
    \caption{\textbf{Kraskov mutual information estimator}. A brief cartoon detailing the basic logic of the Kraskov mutual information estimator \protect\cite{kraskov_estimating_2004}. For each point, a diameter is defined by the distance from that point to it's fourth nearest neighbor. The estimator then counts the number of points within the diameter projected down to the constituent axes, and from that computes the local (pointwise) mutual information \cite{lizier_jidt_2014}. The expected value over all the pointwise values gives the estimated mutual information.}
    \label{fig:kraskov_explainer}
\end{figure}

Typically, information-theoretic measures are computed on discrete probability distributions, where variables can take on a finite number of mutually-exclusive states. For continuous random variables, Shannon defined the differential entropy:
\begin{align}
    \ent(\textbf{X})= - \int d\textbf{x}\text{ }\prob(\textbf{x})\log\prob(\textbf{x}) 
\end{align}
and from this, continuous analogs of all previously introduced measures can be constructed. Unlike discrete probability distributions, however, the problem of estimating $\prob(\textbf{x})$ from data is more involved. If one can assume that the data is multivariate-normal, Gaussian estimators exist and are popular in neuroscience, particularly when using fMRI signals (e.g. \cite{varley_multivariate_2023,luppi_synergistic_2022}). However, for continuous, real-valued data that doesn't satisfy parametric assumptions, a comparatively novel class of estimators based on K-nearest neighbors graphs have been developed. 

A detailed formal treatment is beyond the scope of this manuscript, but the fundamental intuitions are reasonably straightforward. Imagine sampling points from some (potentially high-dimensional) probability manifold with non-trivial structure. Sampled points will, necessarily, cluster more in the high-probability regions, while points in low-probability regions will be more isolated. The distance from a given point to its $k^{\textnormal{th}}$ nearest neighbor can then be used to estimate probabilities. The longer the distance, the lower the local probability in that region. This forms the basis of the Kozachenko-Leonenko non-parametric entropy estimator \cite{kozachenko_sample_1987,delattre_kozachenkoleonenko_2017}, and subsequent derivatives. For a visualization, see Figure~\ref{fig:koz}.

The original estimators, developed by Kozachenko and Leonenko~\cite{kozachenko_sample_1987}, have since been generalized and refined. The most well-known of these is the Kraskov (or KSG) mutual information estimator \cite{kraskov_estimating_2004}, which provides a non-parametric estimator of the coupling between two variables. Visualized in Figure \ref{fig:kraskov_explainer}, for two variables $X$ and $Y$, for each data point, the Kraskov estimator counts the number of points along the marginal $X$ and $Y$ axes that fall between the point and its $k^{\textnormal{th}}$-nearest neighbor. This adaptive approach allows estimates of mutual information, and subsequently, generalizations such as the total correlation, dual total correlation, O-information, and S-information directly from point clouds without the need for discretizing the data or imposing parametric assumptions. For formal details of the non-parametric O-information estimator, see Sec. \ref{sec:estimator}. All measures reported here are implemented in the \texttt{JIDT} package \cite{lizier_jidt_2014}. Distance was defined using the Chebyshev distance for consistency with the TDA analysis using the \texttt{Ripser} package (described below).

\subsection{Topological data analysis and the Rips filtration}

\begin{figure}
    \centering
    \includegraphics[width=\linewidth]{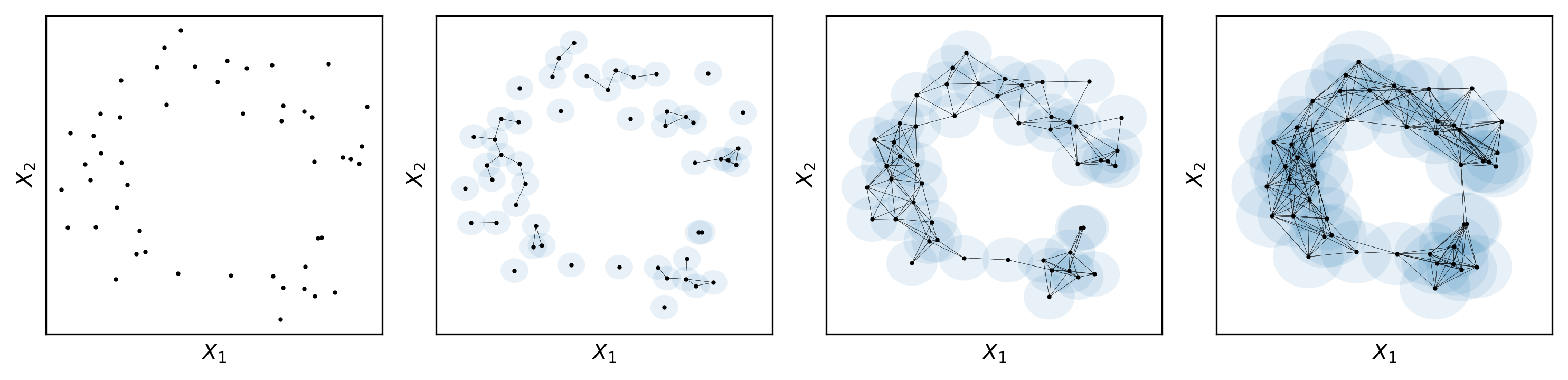}
    \caption{\textbf{Rips filtration}. Consider a two-dimensional point cloud arranged into a rough ring (the same cloud as used in Figure \protect\ref{fig:kraskov_explainer}). Around each point, balls (blue circles) are expanded, and when the radii of two balls intersect, an edge is drawn between the points. When the diameter is low, the simplicial complex is disconnected, comprised on small, tree-like structures. As the balls expand, the simplicial complex becomes denser, and large-scale structures in the data are revealed (in this case, the central void). Eventually, the diameters will be so large that the whole complex will be densely connected and the void will close. }
    \label{fig:rips_explainer}
\end{figure}

The other data-driven approach to higher-order structures is topological data analysis, which is based on algebraic topology; a field of mathematics that combines techniques from abstract algebra and topology to study the properties of high-dimensional structures and spaces. The core intuition behind TDA is identifying ``voids" or ``cavities" in the structure of some high-dimensional point cloud. These ``forbidden" regions correspond to configurations or states that the system under study is restricted from adopting, suggesting some kind of global integration that governs the structure of the part-whole interactions \cite{chazal_introduction_2017,gholizadeh_short_2018}.

The basic building block in TDA is a simplex, which is a $k$-dimensional generalization of a triangle (a 1-simplex is two points connected by a single edge, 2-simplex is a triangle made of three points and three-edges, a 3-simplex is a tetrahedron, and so on). Simplexes can be structured into simplicial complexes, which are collections of intersecting simplexes, with the requirement that any face of the simplicial complex is also a simplex in the complex (i.e. if a tetrahedron is in the complex, then each of it's triangular faces is also part of the complex). This recursive structure provides useful mathematical guarantees that can be leveraged to rigorously analyze quantitative data. 

There are many tools that have been developed to infer the presence of voids in data. One of the most popular is the Vietoris-Rips filtration \cite{edelsbrunner_computational_2010}, which identifies voids by ``growing" a simplicial complex from the point cloud. For a visualization, see Figure \ref{fig:rips_explainer}, but briefly, balls are expanded around each point in the point cloud, and if two balls intersect, an edge is drawn between the points (forming a 1-simplex). As the diameter of the balls expand, increasingly distant points are connected and the simplicial complex becomes ever more densely connected until eventually the diameter of the balls becomes large enough that the point cloud becomes fully connected. 

Voids in the point cloud can be identified by the diameter of the  balls at the moment that they form (birth), and the diameter of the balls at the moment they become ``filled in" (death). This defines the persistence lifetime of the void. From this basic framework, a large number of statistics can be calculated, including the maximum persistence (lifetime of the longest-lived void), the number of unique voids, and more complex derivatives such as the persistence entropy or Betti curves \cite{varley_topological_2021}. Here we focus on the average persistence and the number of unique, three-dimensional voids, as basic, easy-to-understand summary statistics describing the higher-order topology of the cloud. 

\section{Results}

\subsection{Manifolds with known structure}

\begin{figure}
    \centering
    \includegraphics[width=\textwidth]{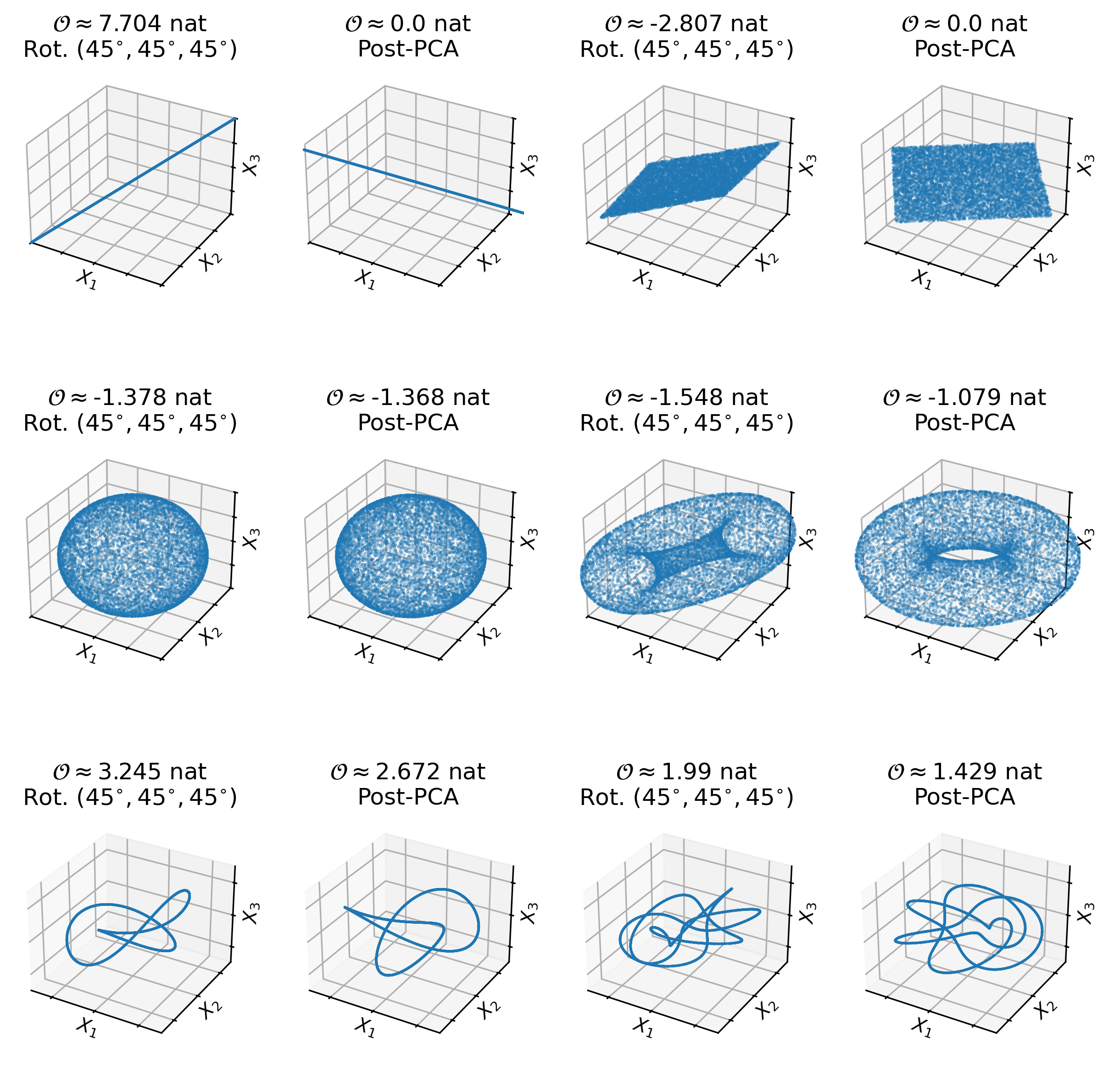}
    \caption{\textbf{O-information of point-clouds with known structure. Top row:} Two point clouds that display only ``contextual" higher-order information. The one-dimensional line, when embedded in a three-dimensional space, is highly redundant, but after rotation with a PCA, all higher-order information is obliterated, as all the information can be represented by one dimension. Similarly, the two-dimensional plane is synergy-dominated when embedded in a three-dimensional space, but also loses its higher-order structure after rotation with PCA. \textbf{Middle row:} Two shapes that have ``intrinsic" synergy associated with three-dimensional cavities. The sphere is perfectly rotationally symmetric, and so no rotation changes the value of the O-information, while the toroid contains a mixture of contextual and higher-order information. \textbf{Bottom row:} Two shapes that contain intrinsic redundancy: a trefoil knot, and its generalization, the $p,q$-knot (\textit{p=5, q=3}). These curves a locally line-like, but cannot be losslessly embeded in a lower-dimensional space. }
    \label{fig:3d_figure}
\end{figure}

We begin by exploring a set of manifolds embedded in a three dimensional space and which have known structure and topological properties. These can serve as intuition-pumps; helping us ground the abstract mathematics of higher-order information with intuitively easy-to-grok examples. 

We begin with the simplest possible three-dimensional point cloud: a one-dimensional line embedded in a three-dimensional space. A one-dimensional random variable, $X_1$ was defined by uniformly sampling 10,000 points on the [-1,1] interval. $X_1$ was then copied two times to construct a three-dimensional random variable, where $X_1=X_2=X_3$. We hypothesized that this system should have very high redundancy, as knowing the state of any $X_i$ immediately resolves all uncertainty about the state of $X_j$ and $X_k$. This was borne out: the KNN-based O-information estimator found a strongly redundancy-dominated structure ($\oinfo=7.704$ nat). Despite this apparent  higher-order information, however, the point cloud itself is fundamentally low dimensional: it is possible to rotate it so that all of the variance falls along the first dimension using PCA. After doing so, the O-information drops to 0. The apparent higher-order information was ``contextual" - it depended on \textit{how} the point cloud was oriented in three-dimensional space, rather than being ``intrinsically" higher-order. This will become a recurring theme.

We can see the same phenomena occur with another very simple manifold: a two-dimensional plane, rotated to be embedded in a three-dimensional space. To construct such a plane, $X_1$ and $X_2$ were independently sampled from the interval [-1,1] (10,000 points each). Then the whole plane was rotated 45$^{\circ}$ degrees along each axis, embedding the plane in a three-dimensional space. This embedded plane had a strongly negative O-information ($\oinfo=-2.819$ nat), however, once again, after rotation using PCA, the O-information dropped to 0. Where does this apparent higher-order synergy come from? It emerges from the fact that, since the plane is a two-dimensional shape rotated into a three-dimensional space, knowing the joint state of $X_1$ and $X_2$ simultaneously is enough to uniquely specify the state of $X_3$. However, knowing either $X_1$ alone or $X_2$ alone resolves very little uncertainty about $X_3$, since for any given value of $X_1$ alone or $X_2$ alone, there are many possible values of $X_3$. When the plane is rotated with PCA, though, it becomes re-embedded in its ``natural" two-dimensional space, and there is no possibility of higher-order structure in a two dimensional space. Once again, we see a distinction between ``contextual" higher-order information, and ``intrinsic" higher-order information. 

Can we construct a point cloud that has ``intrinsic" higher-order information? Yes. Consider a hollow sphere. The sphere itself strongly deviates from global independence (i.e. $\tc(\textbf{X})\gg0$), as it is hollow: the points are constrained to an infinitely thin shell enclosing an empty void. However, if we project the sphere down onto any two-dimensional place, the result is (approximately) a filled circle. We can demonstrate this by sampling 10,000 points from a sphere centered on the origin, and with radius of 1. The resulting O-information is strongly negative ($\oinfo=-1.384$ nat), and crucially, since the sphere is radially symmetric along all axes, putting it through a PCA has absolutely no effect on the O-information at all. It remains -1.384 nat. To test whether it was the presence of the cavity specifically that drove the higher-order synergy (rather than the rotational symmetry), we sampled 10,000 points from a ball with the same dimensions as the sphere, and computed the O-information before and after rotation with PCA.
As expected, the ball had vanishing O-information ($\oinfo=-0.039$ nat, possibly due to bias or variance in the estimator). It did retain the rotational symmetry however: after PCA, the O-information remained the same: ($\oinfo=-0.04$ nat). Based on these results, we argue that it is the topological feature of the hollow sphere (the empty cavity) drives the higher-order synergy, while the geometric rotational symmetry makes it robust to rotation. 

We can further demonstrate this using another shape with a non-trivial topology that includes a three-dimensional cavity: a hollow torus (an inflatable inner-tube). We randomly sampled 10,000 points from the surface of a torus with an minor radius of 0.5 and a major radius of 1, and then rotated the torus $\frac{\pi}{4}$ radians along every axis. The resulting O-information was strongly negative ($\oinfo = -1.554$ nat). When we rotated it with PCA, there was some decrease in synergy, but it remained significantly negative ($\oinfo=-1.096$ nat). Further rotation around the third principal component (which emerges perpendicular to the central ``hole" in the donut) does not change the O-information. 

Like with the hollow sphere, we can assess how the properties of the torus change if we ``fill in" the central, three-dimensional cavity (instead of an inner-tube, picture a solid cake doughnut). If we sample 10,000 points from the interior of a torus with the same dimensions and do the same $\frac{\pi}{4}$ rotation as before, we see $\oinfo=-0.32$ nat; almost an order of magnitude less than the hollow torus. If we rotate the filled torus using PCA, the O-information drops further: $\oinfo=-0.057$ nat, not significantly different from the ball. We can see then that the torus displays a mixture of contextual and intrinsic synergy: it is not completely rotationally symmetric the way the sphere is, but the structure of the internal cavity provides some intrinsic synergy. 

Can we generate a system with intrinsic redundancy? We have seen that the most obvious redundant surface (a straight line) does have significant redundancy, but that it can be rotated away, making it non-intrinsic. We hypothesized that a point cloud with intrinsic redundancy would have to be low-dimensional (locally line-like), but embedded in three-dimensional space in such a way that it could not be losslessly projected down into a lower-dimensional subspace. The natural candidate is a knot: a one-dimensional curve embedded in space in such a way that it never self-intersects. To test this, we generate 10,000 points along a trefoil knot (sometimes also called a triquetra). The trefoil knot is locally one-dimensional, but it requires being embedded into a three-dimensional space to ensure that it never self-intersections. A trefoil knot, rotated by $\frac{\pi}{4}$ radians has a strongly redundant structure ($\oinfo=3.246$ nat). When put through a PCA, the redundancy decreases, however it remains well-above zero ($\oinfo=1.893$ nat). This shows that the simple knot does have intrinsic redundancy. 

The trefoil knot is a special case of a larger family of knots, known as $p,q$-knots (also known as torus knots). These knots provide an interesting counter to the finding that hollow toroids are synergy-dominated, as the $p,q$-knots are all embedded into the surface of a toroid. Consequently, they have superficially similar global structures, but markedly different mathematical structures. To test whether intrinsic redundancy was unique to the trefoil or not, we generated a 5,3-knot, rotated it in the same way as before, and found that it too was significantly redundancy-dominated ($\oinfo=1.96$ nat), and that this redundancy persisted after rotation with PCA ($\oinfo=1.385$ nat). 

These results help us build a foundations of intuitions on how to think about the relationship between topological structure and higher-order information. Several basic features are worth considering. The first is that, while the topology of a point cloud is invariant to how the cloud is rotated, the higher-order information is not. This leads to a distinction that can be made between ``intrinsic" higher-order information, which is that information that is fundamentally tied to the structure of the point cloud regardless of its orientation, and ``contextual" higher-order information, which depends on the specific orientation of the data. The second point is that synergistic information is associated with three-dimensional cavities, or voids in the data. This suggests a natural link between higher-order synergistic information, and topological data analysis. 

\subsection{fMRI data analysis}

\begin{figure}
    \centering
    \includegraphics[width=1.0\linewidth]{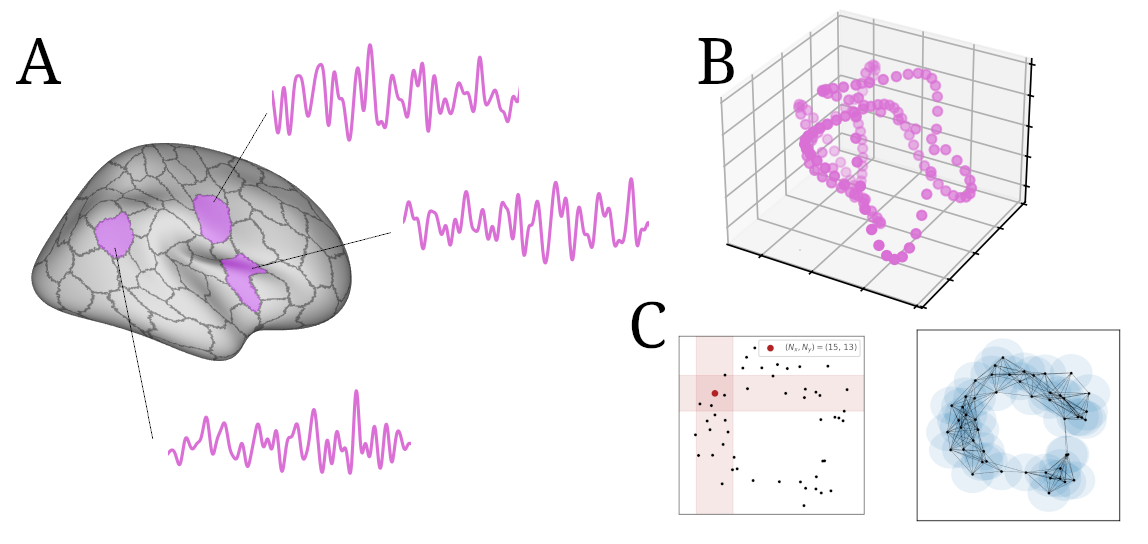}
    \caption{\textbf{Sampling triads and computing measures of higher-order structure. A:} Sets of three brain regions are sampled from the cerebral cortex and their associated BOLD time series are extracted. \textbf{B:} Those time series can be represented as a point cloud embedded in a three-dimensional space (as in \protect\cite{varley_topological_2021}, with each point encoding the joint state of each of the three cortical parcells at time \textit{t}. \textbf{C:} That point-cloud can then be analyzed using persistance homology or multivariate information-theoretic measures.}
    \label{fig:sampling}
\end{figure}

In addition to exploring constructed manifolds with known structure, we also considered how higher-order information and topological data analysis interact in a more naturalistic case. Specifically, we explored multivariate time series data taken from a set of four concatenated resting-state fMRI scans (from a single individual) \cite{van_essen_wu-minn_2013}. Neuroscience as a field has been a testbed for many cutting-edge approaches to studying polyadic interactions between multiple elements, including both information-theoretic and topological approaches. Both information theory and topological data analysis have been found to be informative about diverse cognitive processes, including loss of consciousness \cite{varley_topological_2021,luppi_reduced_2023,luppi_synergistic_2024} and aging \cite{gatica_high-order_2021,rutkowski_mild_2024}.

Briefly, we computed the O-information, total correlation, and dual total correlation using a K-nearest neighbors-based generalization of the Kozachenko-Leonenko entropy estimator \cite{kozachenko_sample_1987,delattre_kozachenkoleonenko_2017}, as implemented by the \texttt{JIDT} package \cite{lizier_jidt_2014}, for all triads of brain regions from a single subject (four concatenated fMRI scans, totaling 4,400 frames). A triad consisted of three brain regions, from which we extracted the associated temporal BOLD data. These three-dimensional time series form a point-cloud in a three-dimensional space, analogous to the point clouds for the spheres, toroids, etc described above (see Figure \ref{fig:sampling}. From this large sample, we selected those triads that showed an O-information significantly greater than, or less than, the expected null O-information computed from an ensemble of autocorrelation-preserving null models (for details, see Materials and Methods). The result was a set of 30,100 significantly redundancy-dominated triads and 6,200 significantly synergy-dominated triads. For each significant triad, we then re-computed the information-theoretic metrics on the point-cloud after it was rotated with principal component analysis, to enable a discussion of ``contextual" versus ``intrinsic" synergy. For the topological data analysis, we computed the average persistence time for three-dimensional cavities and the total number of three-dimensional cavities using the \texttt{Ripser} package for persistence homology \cite{traile_ripserpy_2018}. For both information-theoretic and topological data analyses, distance between points was defined with the Chebyshev metric. 

\subsubsection{Synergy is associated with three-dimensional cavities}

\begin{figure}
    \centering
    \includegraphics[width=\linewidth]{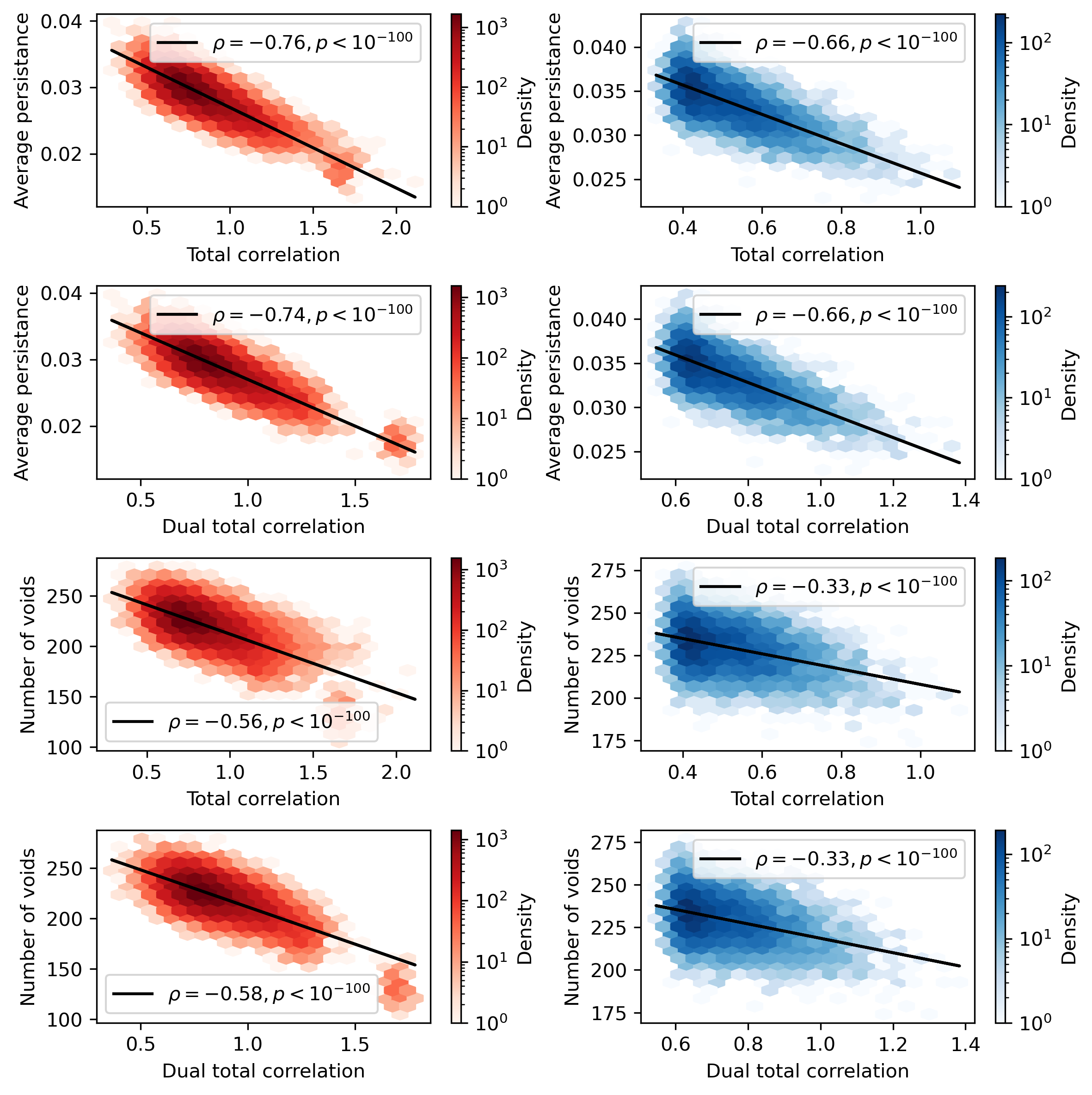}
    \caption{\textbf{Total correlation and dual total corelation are inversely associated with higher-order topological features. Left column:} For redundancy-dominated triads, for both total correlation and dual total correlation, there was a negative relationship between higher-order information and both features of higher-order topology (average persistance and number of voids). \textbf{Right column:} The same pattern (albeit weaker) was also seen in the synergy-dominated triads. This effect was reversed after rotating the pointclouds with principal component analysis (for visualization, see Supplementary Figure \ref{fig:si_2}.}
    \label{fig:dtc_tc}
\end{figure}

We found that for both redundant and synergistic triads, there was a significant negative correlation between average persistence time of three-dimensional voids and both total correlation (redundancy-dominated: $\rho=-0.76$, $p<10^{-100}$, synergy-dominated: $\rho=-0.66$, $p<10^{-100}$) and dual total correlation (redundancy-dominated: $\rho=-0.74$, $p<10^{-100}$, synergy-dominated: $\rho=-0.66$, $p<10^{-100}$). For visualization see Figure \ref{fig:dtc_tc}. Curiously, this effect reverses if the point cloud is rotated using a principal component analysis: the correlations between the information measures and the average persistence become significantly positive (albeit much weaker) for both total correlation (redundancy-dominated: $\rho=0.2$, $p<10^{-100}$, synergy-dominated: $\rho=0.49$, $p<10^{-100}$) and dual total correlation (redundancy-dominated: $\rho=0.22$, $p<10^{-100}$, synergy-dominated: $\rho=0.5$, $p<10^{-100}$).

The same pattern was true when considering the raw number of three-dimensional voids. For both redundant and synergistic triads, there was a significant negative correlation between total number of three-dimensional voids and both total correlation (redundancy-dominated: $\rho=-0.56$, $p<10^{-100}$, synergy-dominated: $\rho=-0.33$, $p<10^{-100}$) and dual total correlation (redundancy-dominated: $\rho=-0.58$, $p<10^{-100}$, synergy-dominated: $\rho=-0.33$, $p<10^{-100}$). Once again this relationship reversed after rotating the point cloud with a PCA transformation: the correlations between the information measures and the number of voids become significantly positive for both total correlation (redundancy-dominated: $\rho=0.1$, $p<10^{-100}$, synergy-dominated: $\rho=0.17$, $p<10^{-100}$) and dual total correlation (redundancy-dominated: $\rho=0.12$, $p<10^{-100}$, synergy-dominated: $\rho=0.18$, $p<10^{-100}$).

Continuing with the distinction introduced above between ``contextual" higher-order information and ``intrinsic" higher-order information, these results show that the relationship between different measures of higher-order structure in data (topological and information-theoretic) is more complex than it might seem at first blush. The contextual total correlation and dual total correlation have a different relationship with the three-dimensional topology of the point cloud than the intrinsic total correlation and dual total correlation. 

\begin{figure}
    \centering
    \includegraphics[width=\linewidth]{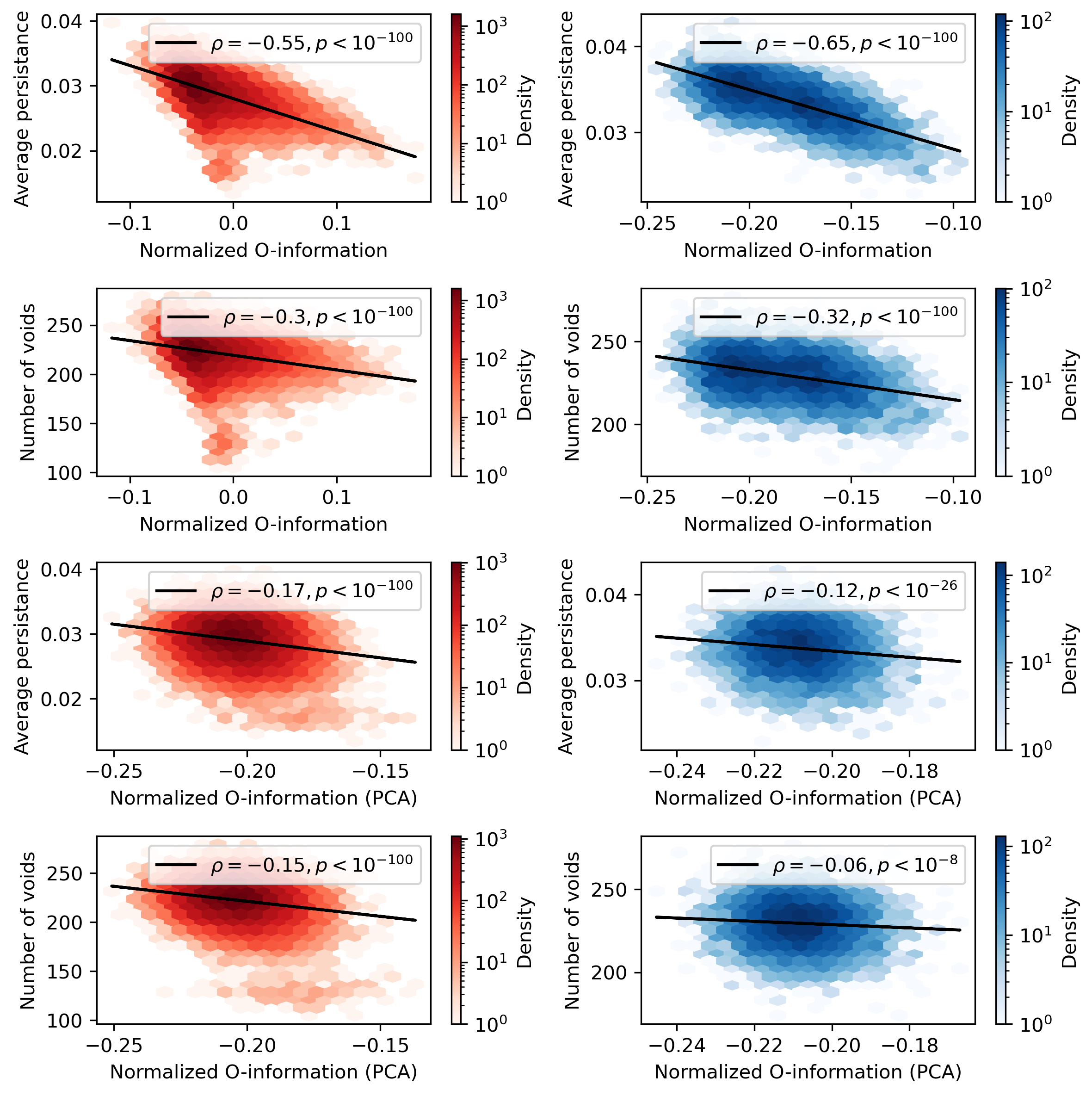}
    \caption{\textbf{Higher-order topological features are inversely correlated with normalized O-information. Right column:} For redundant triads, there were significant negative correlations between the normalized O-information both with (bottom two plots) and without (top two plots) rotation with PCA. \textbf{Left column:} The same pattern was present in the synergistic triads, although after PCA rotation, the negative correlations became much weaker (although still quite significant). Collectively, these results show that the amount of higher-order information in a point cloud is correlated with the presence of higher-order topological features, regardless of how the data is rotated (intrinsic synergy and redundancy).}
    \label{fig:oinfo_topology}
\end{figure}

When we consider the normalized O-information, we find a strong, negative correlation with average persistence of a three dimensional void through the Rips filtration (redundancy-dominated: $-0.55$, $p<10^{-100}$, synergy-dominated: $-0.65$, $p<10^{-100}$). Importantly, unlike with the total correlation and dual total correlation, the direction of the relationship did not change after rotating the point cloud using PCA, although the relationships became considerably weaker (redundancy-dominated: $-0.17$, $p<10^{-100}$, synergy-dominated: $-0.12$, $p<10^{-100}$). Similarly, there is a robust negative correlation between normalized O-information and the number of voids (redundancy-dominated: $\rho=-0.3$, $p<10^{-100}$, synergy-dominated: $\rho=-0.32$, $p<10^{-100}$). As with the average persistence, the direction of the correlation didn't change after rotating the point cloud with PCA, however the relationships became considerably weaker (redundancy-dominated: $\rho=-0.15$, $p<10^{-100}$, synergy-dominated: $\rho=-0.06$, $p<10^{-7}$).

These results show that more negative normalized O-information (i.e. greater synergy dominance) is associated with larger numbers of three-dimensional voids, and longer-lived (i.e. larger) voids as well. Similarly, more positive normalized O-information (i.e. greater redundancy-dominance) is associated with smaller, less numerous cavities. When considered in the context of the results of the analysis of spheres and toroids, these findings suggest that there is a link between the presence of higher-dimensional topological features (specifically cavities) and higher-order information: the presence of one is correlated with greater incidence of the other.

\subsubsection{Low-dimensional manifold analysis fails to represent higher-order structures}

\begin{figure}[]
    \centering
    \includegraphics[width=1\linewidth]{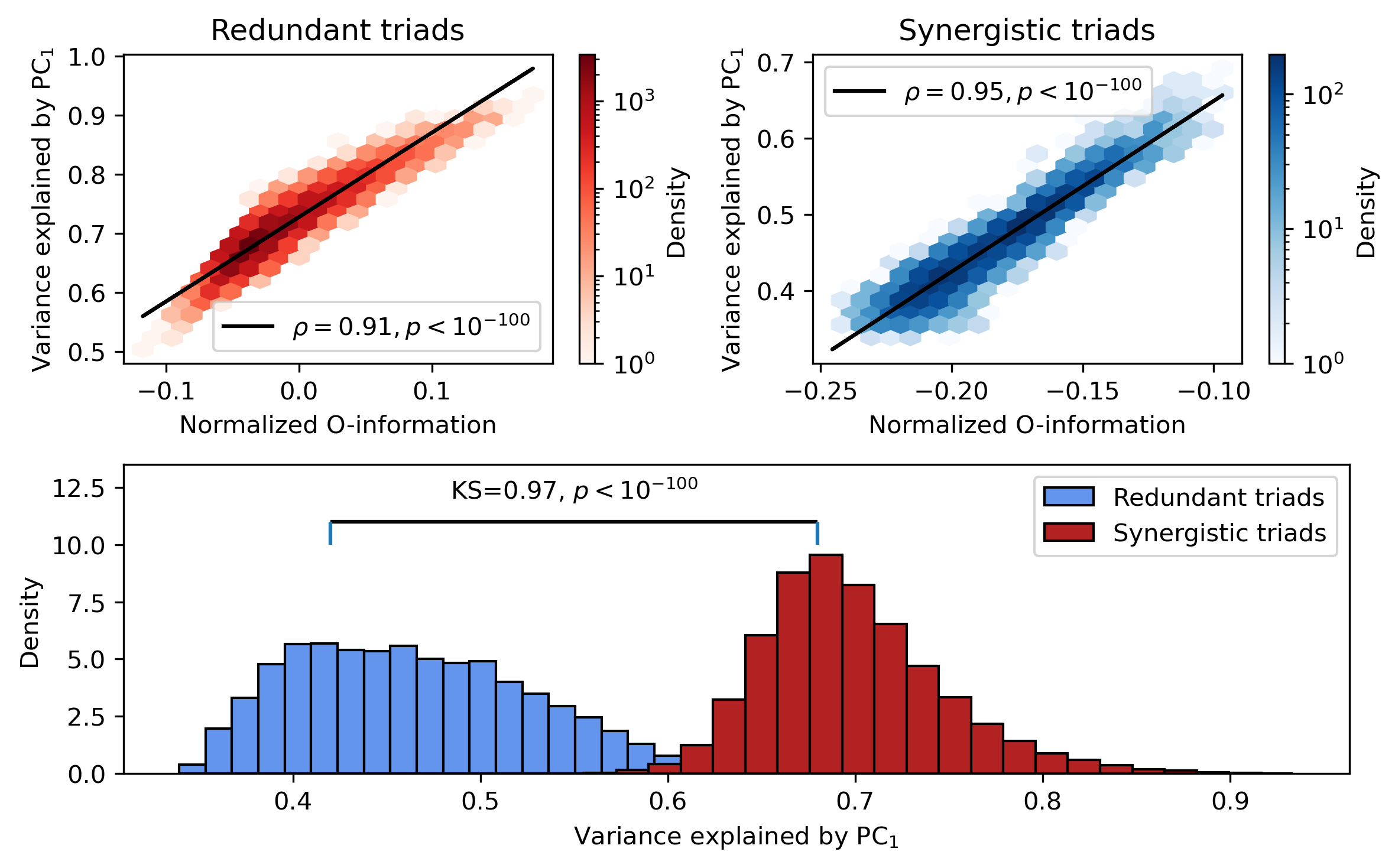}
    \caption{\textbf{Synergy is not captured by low-dimensional manifolds. Top left:} For significantly redundancy-dominated triads, the normalized O-information is strongly correlated with the variance explained by the first principal component, indicating that as the overall amount of redundancy increases, the degree of compressibility does as well. \textbf{Top right:} The same pattern can be seen in synergy-dominated triads as well: a more strongly negative value is associated with decreasing variance explained by the first principal components. Synergistic information is, in a sense, incompressible. Note that, in a totally unstructured system, the variance explained by the first principal component would be $\approx33\%$, which is within the range of observed values for the most synergistic triads, suggesting that the PCA ``sees" these triads as random, despite a strong (higher-order) deviation from independence. \textbf{Bottom:} The distributions of amount of variance explained by the first principal component is also significantly different between redundancy- and synergy-dominated triads, with redundant triads collectively having a greater average compressibility compared to synergistic ones ($KS=0.97$, $p<10^{-100}$). }
    \label{fig:pca_oinfo}
\end{figure}

A very common approach in modern neuroscience is to use manifold learning to extract a low-dimensional representation of highly multivariate data \cite{gallego_neural_2017,gallego_cortical_2018,shine_low-dimensional_2019,langdon_unifying_2023}. It is generally the case that high-dimensional data can be projected down onto a much lower-dimensional manifold that preserves a significant fraction of the total variance. However, it is unclear how high-dimensional information (synergies and redundancies) are represented by these transformations. To test this in the fMRI data, we correlated the proportion of total variance explained by the first principal component of each point cloud against the O-information and topological features. 

For redundant and synergistic triads, there was a significant, positive correlation between the variance explained by the first principal component and both the total correlation (redundancy-dominated: $\rho=0.78$, $p<10^{-100}$, synergy-dominated: $\rho=0.94$, $p<10^{-100}$) and dual total correlation (redundancy-dominated: $\rho=0.67$, $p<10^{-100}$, synergy-dominated: $\rho=0.93$, $p<10^{-100}$). This is as expected, since both total correlation and dual total correlation track different notions of deviation from independence (both measures are zero when every element is independent of every other). 

We found that for both redundancy-dominated and synergy-dominated triads, there was a strong, positive correlation between normalized O-information and the variance explained by the first principal component (redundancy-dominated: $\rho=0.91$, $p<10^{-100}$, synergy-dominated: $\rho=0.95$, $p<10^{-100}$). This result shows that the \textit{more negative} the normalized O-information is (corresponding to comparatively more synergy-dominated structure), the \textit{less} variance is explained by the first principal component. Conversely, triads with strongly positive normalized O-information (corresponding to more redundancy-dominated structure) were more amenable to lossless dimensionality reduction. 

Similar patterns were observed in the TDA measures. For both redundant and synergistic triads, there was a strong negative correlation between the average persistence time of three-dimensional voids and the variance explained by the first PC (redundancy-dominated: $\rho=-0.54$, $p<10^{-100}$, synergy-dominated: $\rho=-0.63$, $p<10^{-100}$). The same pattern was true, albeit weaker, when considering the relationship between the total number of three-dimensional voids at the variance explained by the first PC (redundancy-dominated: $\rho=-0.3$, $p<10^{-100}$, synergy-dominated: $\rho=-0.31$, $p<10^{-100}$). 

Collectively, these results show that PCA-based approaches to manifold learning fail to capture higher-order, synergistic and topologically rich features of the data. Low-dimensional manifolds seem to preferentially represent redundant dependencies (typically in the form of synchronized components) and are blind to synergies. This is consistent with prior work from functional connectivity (FC) analysis, which found that FC networks also preferentially represent redundancies and miss synergies \cite{varley_multivariate_2023,varley_partial_2023}. To reinforce this, we also replicated the functional connectivity results first reported in \cite{varley_multivariate_2023} again with the nearest-neighbors based estimators, rather than the Gaussian estimators (see Supplementary Material figure \ref{fig:si_1}), suggesting that these results are general and not specific to a given class of estimators.

\section{Discussion}

In this paper, we explored the relationships between two different approaches to characterizing higher-order structures in data: multivariate information \cite{varley_information_2024} and topological data analysis \cite{chazal_introduction_2017,gholizadeh_short_2018}. Despite hailing from very different mathematical lineages, we found that topological and information-theoretic approaches to higher-order structures are related. By first analyzing point clouds with known topologies (spheres, planes, hollow toroids, and knots), and then naturalistic data collected from human brain activity, we find evidence of key similarities (and differences) between the two approaches to higher-order structure in multivariate data. 

The most significant finding is that statistical synergy (information in the ``whole" but none of the ``parts") is associated with three-dimensional cavities in the point clouds. In both spheres and toroids, hollow cavities are associated with significantly greater synergy than their solid counterparts. In the fMRI data, both the  number of voids and the average persistance were negatively correlated with normalized O-information, suggesting that both more cavities, and longer-lived cavities appeared in data with more synergy. This suggests that, in a fundamental way, topological data analysis and synergistic information theory are looking at the same kind of underlying structure. Intuitively, one might understand this link by considering Gauss's famous \textit{Theorema Egregium} \cite{gauss_general_2005} that one cannot project a three-dimensional globe onto a two-dimensional surface without deformation: the sphere and a plane are not isometric. This notion of synergistic information present in three-dimensions that is invariably lost when projecting down onto a lower-dimensional space may form the foundation for future methods of estimating synergistic information based on projection distortions. Most information-theoretic treatments of synergy are either coarse (giving a redundancy-synergy balance) like the O-information, or take a redundancy-first approach that implicitly define synergy as ``that information that is left over when all the simpler redundancies are partialed out" \cite{williams_nonnegative_2010}. Truly synergy-first approaches are less well-developed and remain an outstanding problem for the field (for examples, see \cite{rosas_operational_2020,varley_scalable_2024}). 

The finding that redundancy was associated with knots was an unexpected and intriguing one. Recall that $\oinfo(\textbf{X}) > 0 \iff \tc(\textbf{X}) > \dtc(\textbf{X})$. Following the interpretation detailed by Rosas et al., \cite{rosas_quantifying_2019}, we suggest that the structure of the knot can be understood as being dominated by ``collective constraints" (the total correlation) versus ``shared information" (the dual total correlation). Since the knot must be locally one-dimensional at any point, for a given value of $X_1$ in the knot, the set of possible values of $X_2$ and $X_3$ is profoundly constrained by the requirement of local linearity. The collective constraints on the knot are greater than any information shared between the individual $X_i$, leading to a strongly positive O-information. 

We also introduce the distinction between contextual and intrinsic higher-order information. Intrinsic higher-order information is that information which is ``built into" the structure of the point cloud and persists regardless of how the point cloud is rotated. In contrast, contextual higher-order information depends on the specific ways the point cloud loads onto the axes that define the embedding space. Different structures can have either intrinsic synergy or intrinsic redundancy, with intrinsic synergy being associated with rotationally-symmetric three-dimensional cavities and intrinsic redundancy being associated with knots. In the fMRI data, we observed a mixture of intrinsic and contextual higher-order information: generally associations between measures remained significant after rotation with PCA, however the strength of the interactions became much weaker. The significance of these two flavors of higher-order interaction in neural data remains mysterious. 

The finding that manifold learning algorithms like PCA represent redundancies and penalize synergies is consistent with a developing literature exploring how many existing analyses popular in complex systems and computational neuroscience are preferentially biased towards redundancy. Prior work on functional connectivity networks, both theoretical and empirical, has found that statistical networks also reflect higher-order redundancies and are largely blind to synergies \cite{varley_multivariate_2023,varley_partial_2023}. The fact that this also appears to be the case for manifold-learning approaches suggests that synergistic information represents a largely un-explored ``shadow structure" in brain data; missed by currently popular methods like functional connectivity and manifold learning. 

This approach has some limitations that are worth discussing. The first is that no formal proofs relating information theory and topology are provided -- as such these relationships are all correlational in nature, making it more of an exercise in ``experimental mathematics" than a rigorous treatment of the two fields. We conjecture that formal links may be derivable, although this is beyond the current scope of this project. This project is also limited by the computational costs of doing both TDA and non-parametric information theoretic analyses of large datasets. It was not practical to explore triads with more than $\approx$ 1,000 samples, which means that the robustness of the estimates of the underlying manifold may be under-powered. Prior work on synergistic information in fMRI data has been able to leverage hundreds of thousands of samples \cite{varley_multivariate_2023,varley_partial_2023} by using discrete or parametric Gaussian estimators. This is not possible for this paper, and so the results should be interpreted in this context, although we feel that the strength of the observed relationships, as well as consistency between the fMRI results and the shape results, lend credence to our conclusions. 

Finally, this paper is the most recent in a string of studies that connect the idea of information-theoretic synergy to other concepts from mathematics and complex systems. Synergistic information has been linked to ideas in causal inference, with synergies being indicative of causal colliders \cite{varley_synergistic_2024,rosas_characterising_2024}. Synergy has also been linked to chaos in dynamical systems: Boolean networks evolved for highly synergistic dynamics are generally highly chaotic, displaying classic features such as sensitivity to perturbation, long transients, and high-entropy dynamics \cite{varley_evolving_2024}. This study continues in this vein by linking synergistic information to higher-order topological features. 

\section{Methods}

\subsection{Nearest-neighbor estimators for the O-information}
\label{sec:estimator}

[Work in progress]

One of the key methodological requirements for this study is a reliable estimator of high-order interactions in high-dimensional, non-linear distributions. Common methods such as Gaussian~\cite{lizier_jidt_2014} or kernel~\cite{marinazzo2008kernel} estimators are not suitable for this task (Gaussian methods because they are linear, kernel methods because they are very sensitive to hyperparameter choices) -- leaving nearest-neighbor estimators as the natural choice.

Following the seminal work of Kraskov et al.~\cite{kraskov_estimating_2004}, the nearest-neighbor (NN) mutual information estimator consists primarily of two steps: \emph{i}) a neighbor search, to find the distance from each point to its $k^{\textnormal{th}}$ nearest neighbor; and \emph{ii}) a range search, to count the number of points within this distance in each dimension (see Figure \ref{fig:kraskov_explainer} for visualization). These counts are then used to compute the pointwise mutual information, as outlined in Section~\ref{sec:knn} and described in detail in the original paper~\cite{kraskov_estimating_2004}.

A naive approach to estimating O-information with NN methods could be, for example, to estimate each mutual information separately (using the Kozachenko-Leonenko~\cite{kozachenko_sample_1987} or Kraskov~\cite{kraskov_estimating_2004} algorithms) and then adding these to form the O-information. Unfortunately, this would result in a very high bias due to the neighbor searches in spaces with different dimensionality (as already pointed out by Kraskov et al.~\cite{kraskov_estimating_2004}). Our goal, then, is to formulate a nearest-neighbor O-information estimator that only involves one neighbor search, so bias is kept to a minimum.

This can be achieved thanks to the extension to Kraskov et al.'s ``Algorithm 1"~\cite{kraskov_estimating_2004} for entropy combinations provided by G\'omez-Herrero et al.~\cite{gomez-herrero_assessing_2015}. To apply this estimator, we begin from the expression of the O-information in terms of marginal entropies~\cite[Def.~1]{rosas_quantifying_2019}:
\begin{align}
    \oinfo(\textbf{X}) = (N-2)\ent(\textbf{X}) - \sum_{i=1}^{N}\left[\ent(X_i) - \ent(\textbf{X}^{-i}) \right] ~ .
\end{align}
Through basic arithmetic, this equation can be mapped to Eq.~(1) of Ref.~\cite{gomez-herrero_assessing_2015}, confirming that the O-information is indeed an entropy combination (scaled by a multiplicative factor). Applying this to Eq.~(3) of \cite{gomez-herrero_assessing_2015}, we obtain the following O-information NN estimator:
\begin{align}
    \hat\oinfo(\textbf{X}) = (2-N)\left[F(k) + \sum_i^N\frac{1}{N-2}\left\langle F(k_i(t)) - F(k^{-i}(t))\right\rangle_t \right] ~ ,
    \label{eq:knn_oinfo}
\end{align}
where $F(k) = \psi(k) - \psi(N)$, $\psi$ is the digamma function, and $\langle\cdot\rangle_t$ denotes a sample average across samples indexed by $t$. Denoting by $\varepsilon(t)$ the distance between point $t$ and its $k^{\textnormal{th}}$ nearest neighbor in the joint space $\textbf{X}$, $k_i(t)$ denotes the number of points within distance $\varepsilon(t)$ of point $t$ in the marginal space $X_i$ -- and similarly for $k^{-i}(t)$ in the marginal space $\textbf{X}^{-i}$.

Therefore, the formulation in Eq.~\eqref{eq:knn_oinfo} allows us to estimate O-information using $2N$ range searches ($N$ for the one-dimensional marginals $X_i$ and $N$ for the $(N-1)$-dimensional marginals $\textbf{X}^{-i}$) and, crucially, only one neighbor search -- providing a much less biased estimator of the O-information.

The resulting algorithm is published as part of the JIDT toolbox~\cite{lizier_jidt_2014} and is publicly available at \url{github.com/jlizier/jidt}.

\subsection{Significance-testing O-information}

A confound of any study of multivariate neural time series is the presence of autocorrelation, which can complicate the problem of inference by artificially inflating the apparent dependence between two actually-uncorrelated variables \cite{cliff_exact_2020}. To ensure that we were only analyzing triads with genuine higher-order interactions, we employed a circular shift-based null hypothesis significance test to filter out triads with an O-information likely to be attributable to first-order autocorrelation and not true higher-order dependency. Briefly: we first computed the O-information for each of the 1,313,400 unique triads. We then re-computed the O-information for each of the triads, after circular-shifting each time series so that it came from a different scan (since four scans were appended to infer the joint distribution. This means that each of the three channels were recorded at different points in the scan, sometimes hours apart, and any O-information is attributable just to the autocorrelation (which is preserved by the circular shift). 

From this set of 1,313,400 nulls, we built a distribution against which to test the empirical, un-perturbed O-informations. If a triad had an empirical O-information lower than three standard deviations from the mean of the null distribution, it was said to be significantly synergistic, and if it was greater than three standard deviations from the mean of the null it was said to be significantly redundant. The result was 30,100 significantly redundancy-dominated triads and 6,200 significantly synergy-dominated triads. 

\subsection{Topological data analysis}

All topological data analysis was done using the Ripser package \cite{traile_ripserpy_2018}, on a pre-computed Chebyshev distance matrix with the maximum cohomology dimension set to 2. Since the persistence homology computation grows unmanageably with the number of points, we randomly sampled 1,100 frames from the 4,400 concatenated BOLD time series (equivalent to a single scan) to minimize excessive computation while still making sure that each of the four constituent scans contributed. All persistence statistics were computed from the resulting persistence diagram.   

\subsection{Data collection and preprocessing}

The data used here has been previously described in a number of prior papers \cite{pope_modular_2021,varley_multivariate_2023,varley_partial_2023}. We will briefly reproduce a high-level overview from \cite{varley_partial_2023}. The data used in this study was taken from a set of 100 unrelated subjects included in the Human Connectome Project (HCP) \cite{van_essen_wu-minn_2013}. All subjects gave informed consent to protocols approved by the Washington University Institutional Review Board. Data was collected with a Siemens 3T Connectom Skyra using a head coil with 32 channels. Functional data analysed here was acquired during resting state with a gradient-echo echo-planar imaging (EPI) sequence. Collection occurred over four scans on two separate days (scan duration: 14:33 min; eyes open). The main acquisition parameters included TR = 720 ms, TE = 33.1 ms, flip angle of 52$^{circ}$, 2 mm isotropic voxel resolution, and a multiband factor of 8. Resting state data was mapped to a 200-node parcellation scheme \cite{schaefer_local-global_2018} covering the entire cerebral cortex. Considerations for subject inclusion were established before the study and are as follows. The mean and mean absolute deviation of the relative root mean square (RMS) motion throughout any of the four resting scans were calculated. Subjects that exceeded 1.5 times the interquartile range in the adverse direction for two or more measures they were excluded. This resulted in the exclusion of four subjects, and an additional subject due to a software error during diffusion MRI processing.

The minimal preprocessing of HCP rs-fMRI data can be found described in detail in \cite{glasser_minimal_2013} Five main steps were followed: 1) susceptibility, distortion, and motion correction; 2) registration to subject-specific T1-weighted data; 3) bias and intensity normalization; 4) projection onto the 32k fs LR mesh; and 5) alignment to common space with a multimodal surface registration. This pipeline produced an ICA+FIX time series in the CIFTI grayordinate coordinate system. We included two additional preprocessing steps: 6) global signal regression and 7) detrending and band pass filtering (0.008 to 0.08 Hz) \cite{parkes_evaluation_2018}. We discarded the first and last 50 frames of each time series after confound regression and filtering to produce final scans with length 13.2 min (1,100 frames)

\subsubsection*{Acknowledgments}
TFV would like to thank Joe Lizier for assistance with the JIDT package. The authors received no special funding for this work. 

\subsubsection*{Author contributions}
TFV conceptualized the project, performed analyses, and wrote the manuscript. PM wrote analytic code, contributed to the manuscript, and gave feedback. JB provided supervision and feedback throughout the project. 

\subsubsection*{Competing interests}
The authors have no competing interests to declare. 

\subsubsection*{Ethics declaration}
No new data was collected for the purposes of this study. All HCP participants participants provided written informed consent to data collection protocols approved by the Ethics Committee of the Montreal Neurological Institute and Hospital.

\subsubsection{Code and data availability}
Code for this paper will be published as supplementary material in the final version. The HCP fMRI scans are available from \url{https://www.humanconnectome.org}.

\bibliography{main}

\newpage

\section*{Supplementary Material}

\subsection*{Network structure and normalized O-information}

\begin{figure}[h!]
    \centering
    \includegraphics[width=\linewidth]{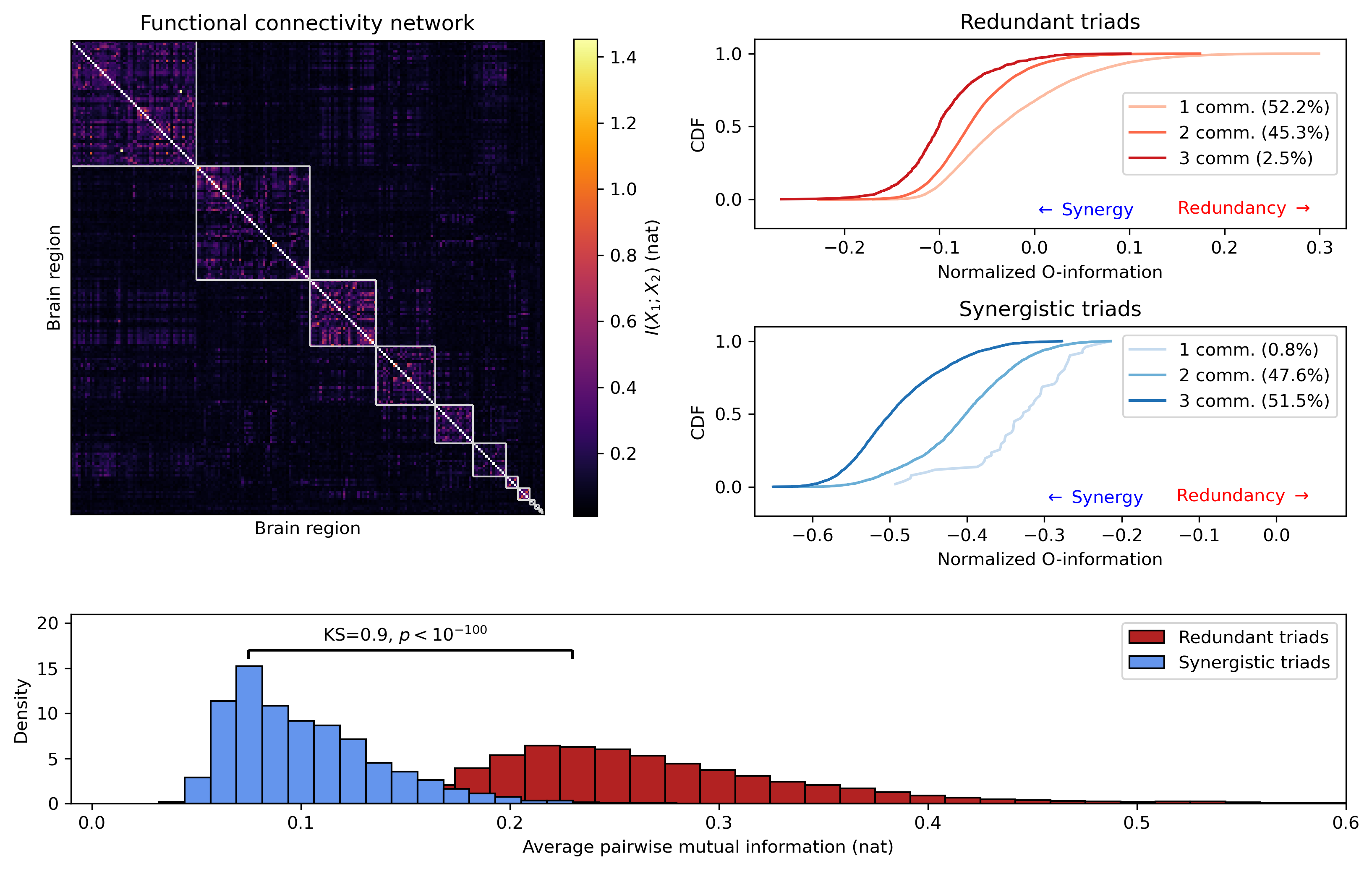}
    \caption{\textbf{The relationship between Kraskov O-information and bivariate functional connectivity structure. Top left:} The functional connectivity network constructed from an fMRI scan of a single person. Each edge gives the non-parametric, mutual information (estimated with the KSG algorithm \protect\cite{kraskov_estimating_2004}) between two brain regions. The matrix has been organized into functional clusters with higher within-module density and low between module density using multi-resolution consensus clustering \protect\cite{jeub_multiresolution_2018}. \textbf{Top right:} We separated each of the triads in the redundancy- and synergy-dominated groups into triads that were contained within one functional community, spanned two functional communities, or had each element in three distinct communities. We then plotted cumulative distribution functions on the normalized O-information for each. We can see that, for redundant triads, those that spanned three communities had the weakest redundancy, while those that sat all within one community had the greatest redundancy. Similarly, for synergistic triads, those that spanned three communities had the greatest synergy, while those that all sat within one community (which were extremely rare, comprising les than 1\% of synergistic triads) had the weakest synergy. \textbf{Bottom:} The distribution of pairwise mutual information (functional connectivity) for redundancy- and synergy-dominated triads was significantly different ($KS=0.9$, $p<10^{-100}$). \newline Collectively, these results replicate findings first reported by \protect\cite{varley_partial_2023}, using the non-parametric, continuous estimators as opposed to the previously used discrete estimator. }
    \label{fig:si_1}
\end{figure}

\newpage 

\subsection*{PCA inverts the relationship between total correlation, dual total correlation, and topological features}

\begin{figure}[!h]
    \centering
    \includegraphics[width=\linewidth]{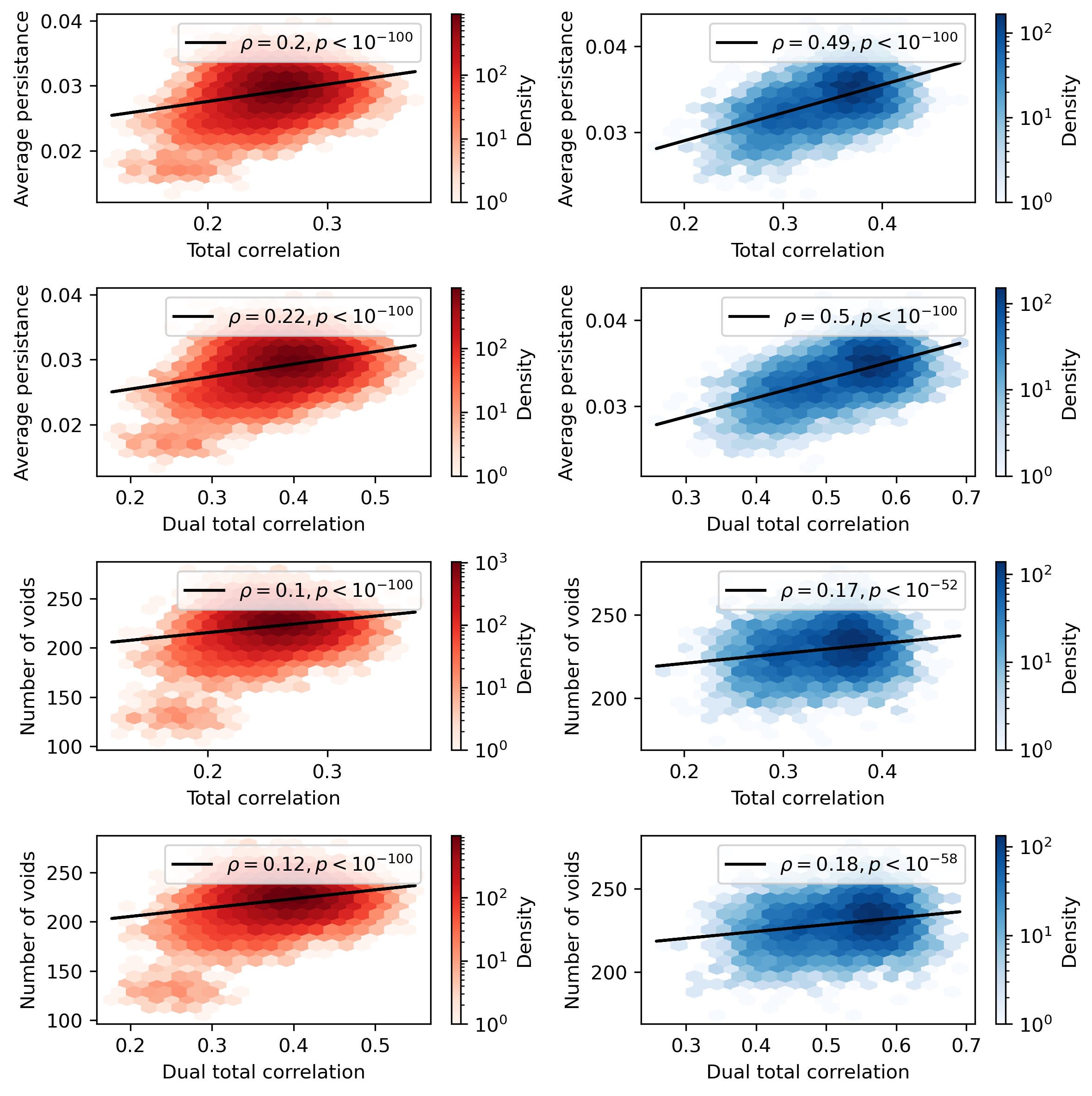}
    \caption{\textbf{Rotating point clouds inverts the relationship between TC, DTC, and higher-order topological features.} The same plot as shown in Fig. \protect\ref{fig:dtc_tc}, however the point-clouds have been rotated with PCA. Note that the direction of the relationship between multivariate information and topology reverses.}
    \label{fig:si_2}
\end{figure}

\newpage 

\subsection*{Logical XOR gate analysis}

\begin{table}[h!]
\begin{center}
\begin{tabular}{|c|c|c|c|}
    \toprule
    $\mathbb{P}(\textbf{X})$ & $X_1$ & $X_2$ & $X_3$ \\
    \midrule
    $1/4$ & 0 & 0 & 0 \\
    $1/4$ & 0 & 1 & 1 \\
    $1/4$ & 1 & 0 & 1 \\
    $1/4$ & 1 & 1 & 0 \\
    \bottomrule
\end{tabular}
\caption{The lookup table for a stochastic logical XOR gate.}
\end{center}
\end{table}

The logical XOR date is a classic example of an irreducibly higher-order structure in a multivariate system. For any pair of elements, the pairwise mutual information $I(X_i ; X_j) = 0$ bit. However, the joint mutual information $I(\{X_i,X_j\}; X_k) = 1$ bit: there is information in the ``whole" trivariate system that is not reducible to individual pairwise interactions. A ``functional connectivity" network model of an XOR gate would be indistinguishable from a functional connectivity network of three independent processes. It is only when higher-order synergies are accounted for that the dependency structure becomes visible. 
\end{document}